\documentclass{article}      
\usepackage{adjustbox}
\usepackage{graphicx}
\usepackage{epstopdf}

\title{The baryon spectrum and the hypercentral Constituent Quark Model  }  
\author{
M.M. Giannini\\
Dipartimento di Fisica dell'Universit\`a di Genova\\
and \\
I.N.F.N., Sezione di Genova\\
E. Santopinto\\
I.N.F.N., Sezione di Genova\
}
\date{}
\begin{document}             

\maketitle         

\begin{abstract}
The description of the baryon spectrum is performed using the hypercentral Consituent Quark Model (hCQM), mainly in comparison with  the harmonic oscillator (h.o.). Recentlly many new states, at various levels of  confidence have been observed, leading to a softening of the missing resonance problem in the case of positive parity states. However, the number of negative states is higher that predicted by the commonly used h.o. scheme and therefore one  is forced to take into account also the higher energy shells, which contain an overall number of states much greater than the observed one. 
It is shown that, thanks to the peculiar level scheme of the hCQM, the recently observed negative parity states can be considered as belonging  to the lower shells, keeping the missing resonance problem within more acceptable limits.
\end{abstract}

\section{Introduction}

The recent editions of the PDG report \cite{pdg12,pdg14} some new non strange resonances  at various levels of confidence. The observed states are compared with the spectrum predicted by the Constituent Quark Model \cite{QM} and also with the most recent results obtained with Lattice QCD calculations \cite{edw}.

It is well known that quark models predict a relevant number of states which are not observed, leading to the so called problem of the missing resonances. Up to few years ago,   the non strange states have been mainly observed in the pion channel, while, according to ref. \cite{ki}, many of the unobserved resonances are expected to be preferably seen in photo- or electro-production of mesons. This idea is supported by the recents analysis of the photon channel \cite{anis}, but the problem is still present, even if it is softened by the new findings \cite{pdg12,pdg14}. 
It is interesting to observe that the number of the states identified in  the Lattice QCD calculations \cite{edw} is much larger than the ones predicted by any CQM.

There are various Constituent Quark Models (CQMs), which have been applied with success to the description of the nucleon internal structure, namely  the Isgur-Karl \cite{ik}, the relativized Capstick-Isgur
\cite{ci}, the algebraic  \cite{bil}, the hypercentral 
 \cite{pl}, the chiral Goldstone Boson Exchange 
\cite{olof,ple} and the Bonn instanton models \cite{bn}. 

When  the problem of the missing resonances is discussed, quite often reference is still made to an underlying harmonic oscillator (h.o.) model. The reason is of course because the model is simple and analytically solvable and therefore one forgets that  the h.o. spectrum is too rigid to be realistic. However, if the classification of the states provided by the h.o. is taken seriously, it is difficult to include in the picture some of the newly observed resonances, namely the new negative parity states.

In the following we shall show that this difficulty can be avoided if the hCQM \cite{pl} is taken as a reference framework. Such model has been applied to the description of various quantities of physical interest: the spectrum \cite{pl,sig,sig2,iso}, the photocouplings \cite{aie},  the helicity amplitudes for the excitation of the nucleon resonances \cite{aie2,mds2,sg} and the elastic form factors \cite{mds,rap,ff_07,ff_10} (for a review see ref. \cite{chin}), it is not analytical but the pattern of the spectrum, although different from the h.o. one, can be formulated in a simple way, similarly to the h.o. case.

\section{The description of the spectrum}

The structure of the spectrum provided by the major part of the CQMs is the same as the one predicted by the h.o. model of the Isgur-Karl type. One can assume that the dominant part of the quark interaction is spin-independent \cite{deru} and the three quark states can be classified according to the SU(6) group. The three quark wave function $\Psi_{3q}$ can be factorized as follows
 \begin{equation}
\Psi_{3q}~=~\theta_{colour} ~\Phi_{SU(6)}~\psi_{space}
\label{3q_2}
\end{equation}
$\theta_{colour}$ must be a colour singlet and is completely antisymmetric for the exchange of any quark pair. Then the remaining part of the wave function must be completely symmetric and the space and SU(6) parts share the same permutational symmetry.

Since quarks form the six dimensional representation of SU(6), 
the baryon states belong to the SU(6) multiplets obtained by the decomposition of the product 
\begin{equation}
6 \otimes 6  \otimes 6= 20_A  \oplus 70_M  \oplus 70_M \oplus 56_S 
\end{equation}
the symmetry type A,  M, M, S being explicitly indicated.

\begin{table}[t]
\caption[]{The non strange three-quark states belonging to  the h.o. shells up to $N_{h.o.}=2$. The notation for the baryon resonances is $X J^P$, where X=N or $\Delta$, denotes the type of the nonstrange baryon and $J^P$ is the spin-parity of the state. The asterisk means the presence of a radially excited wave function. }
\label{ho}
\vspace{15pt}
\begin{center}
\begin{tabular}{|c|c|c|c|c|c|}
\hline
$N_{h.o.}$ & $(d,L^P)$ & $^2\underline{8}$ & $^48$ & $^210$ & $^410$ \\
\hline
\hline
& & & & & \\
 0 & $(56,0^+)$ & $N 1/2^+$ &  &  &  $\Delta 3/2^+$ \\
 & & & & & \\
\hline
\hline
& & & & & \\
 1 & $(70,1^-)$ & $N 1/2^-$ & $N 1/2^- $ &$\Delta 1/2^-$  &  \\
  & & $N 3/2^-$ & $N 3/2^-$&$\Delta 3/2^-$ &\\
 & & &$N 5/2^-$ & & \\
  & & & & & \\

\hline
\hline
& & & & & \\
 2 & $(56^*,0^+)$ & $N 1/2^+$ &  &  &  $\Delta 3/2^+$ \\
 & & & & & \\
\hline
& & & & & \\
 2 & $(70,0^+)$ & $N 1/2^+$ & $N 3/2^+$ & $\Delta 1/2^+$ &   \\
 & & & & & \\
\hline
& & & & & \\
 2 & $(56,2^+)$ & $N 3/2^+$ &  &  & $\Delta 1/2^+$  \\
 & &$N 5/2^+$ & & & $\Delta 3/2^+$\\
  & & & & & $\Delta 5/2^+$\\
   & & & && $\Delta 7/2^+$ \\
    & & & & & \\
\hline
  &  &  & & &   \\
 2 & $(70,2^+)$ & $N 3/2^+$ & $N 1/2^+$ & $\Delta 3/2^+$ &   \\
  &  &  $N 5/2^+$& $N 3/2^+$ & $\Delta 5/2^+$&   \\
  &  &  & $N 5/2^+$& &   \\
  &  &  & $N 7/2^+$& &   \\
  &  &  & & &   \\
\hline
&  &  & & &   \\
 2 & $(20,1^+)$ & $N \frac{1}{2}^+$ &  & &   \\
 &  & $N \frac{3}{2}^+$ & & &   \\
 &  &  & & &   \\
\hline
\end{tabular}
\end{center}
\end{table}

The $SU(6)$ representations can be decomposed according to their 
 spin and flavour content
 \begin{equation}
20~=~^4 \underline{1}~+~^2 \underline{8} 
\end{equation}
\begin{equation}
56~=~^2 \underline{8}~+ ~^4 \underline{10}  
\end{equation}
\begin{equation}
70~=~^2 \underline{1}~+~^2 \underline{8}~+~^4 \underline{8}~+~^2
\underline{10} 
\\
\end{equation}
The suffixes in the r.h.s. denote the multiplicity $2S+1$ of the $3q$ spin
states
and the underlined numbers are the dimensions of the $SU(3)$
representations.  

This pattern is quite general and determines the type and number of three quark states that can be built, taking into account the possible values of the total orbital angular momentum and the total spin of  the three quarks.   The construction of the states must of course be performed in order to respect the  symmetry property required by the Pauli principle.

Referring to the h.o. levels, it is customary to consider only the states with $N_{h.o.} \leq 2$, that is within the first three shells. The states with $N_{h.o.}=3$ have been considered in order to account for all the negative parity states, including the higher energy ones \cite{fors}, while the $N_{h.o.}=4$ have been used in order to test the quark model predictions for the $N-\Delta$ quadrupole transition \cite{quad}. The introduction of the higher shell is certainly needed for the description of the higher part of the spectrum, where, however, the number of the observed resonances is very small, in contrast with the very large number of states contained in the shells with $N_{h.o.}=3,4$. One should say that the opening of decay channels increases as soon as the energy becomes higher and the resonances are expected to be very large and strongly overlapping, making it very difficult to observe them, apart from some states with high spin values.

\begin{table}[h]
\caption[]{The number of h.o. states with $N_{h.o.} \leq 2$, reporting  the positive and negative parity N and $\Delta$ states separately.  In the third column, also the states predicted by the hCQM, described later on in the text, are given. In the last six columns, the number of states listed in the 2010 \cite{pdg10} and 2014 \cite{pdg14} PDG editions  are given separately according to their star assignments.}
\label{comp}
\vspace{0.4cm}
\begin{center}
\begin{adjustbox}{max width=\textwidth}
\begin{tabular}{|c|c|c||c|c||c|c||c|c|}
\hline
& & &  & & & & &  \\
 & h.o. & hCQM & PDG10  &PDG14  &PDG10   & PDG14 &PDG10   & PDG14   \\
 & &  &  4* 3* & 4* 3* & 4* 3* 2*  & 4* 3* 2* & all stars  & all stars \\
&  & & &  & & & & \\
\hline
& &  & & & & & & \\
$N^+$ & 14  & 15 & 5 & 6 & 8 & 10 &  9 & 12  \\
& & &  & & & & & \\
 $N^-$ &5  & 10 & 5 & 6 & 7  & 9 & 8 & 9 \\
& & &  & & & & &  \\
\hline
& & &  & & & & &  \\
 N tot & 19 & 25  & 10 &12 &15  & 19 & 17 & 21\\
& & &  & & & & &  \\
\hline
& & &  & & & & &  \\
$ \Delta^+$& 9 &  10&  6 &  6 & 7 & 7 & 8  & 8 \\
& &  & & & & & & \\
$ \Delta^-$& 2 & 4 & 2 & 2 & 3 & 4 & 4 & 4 \\
& & &  &  & & & &\\
\hline
& & &  & & & & &  \\
 $\Delta tot$ & 11 & 14  & 8 &8 & 10  & 11 & 12 & 12\\
& & &  & & & & &  \\
\hline
\hline
& &  & &  & & & & \\
Total& 30 & 39 & 18 &  20 & 25 & 30 & 29 & 34  \\
& &  & &  & & & & \\
\hline
\end{tabular}
\end{adjustbox}
\end{center}
\end{table}

We limit then ourselves to the first three h.o. shells and the non strange three quark states which are obtained in this way are shown in Table \ref{ho}. In order  to determine which resonances can be arranged in these SU(6) multiplets, we report in Table \ref{comp}, for each type of state, the number of the h.o. states and those reported in the 2010 and 2014 PDG editions. The states predicted by the hCQM, described later, are also given. 

From Table \ref{comp} we have excluded those states which cannot be built from h.o. states with $N_{h.o.} \leq 2$; in this respect we observe that the maximal spin values  of the states in Table \ref{ho} are $7/2^+$, $5/2^-$ for N-  and $7/2^+$, $3/2^-$ for $\Delta$-type states, respectively. 
This means in particular that the  $3*$ $\Delta(1930) 5/2^-$ belongs to the $N_{h.o.}=3$ shell. In fact, one could in principle obtain a spin 5/2 state starting from a total orbital angular momentum L=1 and total spin $S=3/2$, however, since the spin and isospin parts of the wave functions are completely symmetrical, the space part should also be completely symmetrical, while the L=1 state has mixed symmetry. 

Furthermore, we have excluded from the table also the $2 *$ states $N(2570)	5/2^-$ and  $N(2300) 1/2^+$, having an energy two high in comparison with that of the resonances belonging to the first three shells. It should be noted that the four $1 *$ resonances quoted in Table \ref{comp} are poorly known, for the states $\Delta (1750) 1/2^+$ and $\Delta(2150) 1/2^-$ there is not even the indication of the mass interval \cite{pdg14}; the latter is also excluded from Table \ref{comp}, since, as it will be clear later on, it probably belongs to a higher shell.

Looking at Table \ref{comp} one sees that the situation for positive and negative parity states is quite different. For simplicity we refer to the resonances reported in the 2014 PDG edition \cite{pdg14}.

The number of positive parity states with 4 and 3 stars is lower than the h.o. levels, giving rise to the well known problem of the missing resonances. The problem has been notably softened  by the newly observed states \cite{pdg12,pdg14}, moreover,  if one takes into account also the 2 star states, the discrepancy becomes less meaningful. 

The reverse happens for the negative parity states. Up to 2010, the seven observed states were just those predicted by the h.o. classification, but now there is a new three star state $N(1875) 3/2^-$ \cite{pdg12,pdg14}, which has an energy comparable with the other negative parity states and can be hardly thought to belong to the $N_{h.o.}=3$ shell. If one considers also the two star states, the number of the observed negative parity resonances is 13, while  the theoretical ones are 7. 

A  possible assignment of the resonances, including the new ones,  to the SU(6) configurations is given in Table \ref{ho2}. 

In the $N_{h.o.}=2$ shell, six states are missing. Some may be given by one star resonances, namely the $N(2100) 1/2+$ and $N(2040) 3/2+ $ states, probably fitting into the $(20,1^+)$ configuration and the $\Delta(1750) 1/2^+$, which could be included in the $(70,0^+)$ configuration. In this way, what are really missing are two N- and one $\Delta$ states, all with $J^P= 3/2^+$.
All the remaining states reported in the PDG \cite{pdg14} have spin-parity values possible only considering the $N_{h.o.}=3,4$ shells. These states have for the major part energies higher than the states reported in Table \ref{ho2}, with the exception of the already mentioned three star $\Delta(1930) 5/2^-$.

\begin{table}[h]
\caption[]{The non strange three-quark states belonging to  the h.o. shells up to $N_{h.o.}=3$. For the baryon resonances we use the notation introduced recently by the PDG \cite{pdg10}. The resonances with four and three stars are reported in bold face, the remaining are the two star states. The assignments of the $N_{h.o.} \leq 2 $ resonances to the SU(6) configurations coincide with those of ref. \cite{ik}. For the $N_{h.o.}=3$, only one configuration is reported. }
\label{ho2}
\vspace{15pt}
\begin{center}
\begin{adjustbox}{max width=\textwidth}
\begin{tabular}{|c|c|c|c|c|c|}
\hline
$N_{h.o.}$ & $(d,L^P)$ & $^2\underline{8}$ & $^4\underline{8}$ & $^2\underline{10}$ & $^4\underline{10}$ \\
\hline
\hline
 0 & $(56,0^+)$ & $\bf{N(939) 1/2^+}$ &  &  &  $\bf{\Delta(1232) 3/2^+}$ \\
\hline
\hline
 1 & $(70,1^-)$ & $\bf{N(1535) 1/2^-}$ & $\bf{N(1650) 1/2^-} $ &$\bf{\Delta(1620)}1/2^-$  &  \\
  &  & $\bf{N(1520) 3/2^-}$ & $\bf{N(1700) 3/2^-} $ &$\bf{\Delta(1700)} 3/2^-$  &  \\
   &  &  & $\bf{N(1675) 5/2^-} $ &  &  \\
\hline
\hline
 2 & $(56^*,0^+)$ & $\bf{N(1440) 1/2^+}$ &  &  &  $\bf{\Delta(1600) 3/2^+}$ \\
\hline
 2 & $(70,0^+)$ & $\bf{N(1710) 1/2+}$ & $\bf{N(1900) 3/2^+}$ & $\Delta 1/2^+$ &   \\
\hline
 2 & $(56,2^+)$ & $\bf{N(1720) 3/2^+}$ &  &  & $\bf{\Delta(1910) 1/2^+}$  \\
  &  & $\bf{N(1680) 5/2^+ }$ &  &  & $\bf{\Delta(1920) 3/2^+} $  \\
  &  &  &  &  & $\bf{\Delta(1905) 5/2^+}$  \\
  &  &  &  &  & $\bf{\Delta(1950) 7/2^+}$  \\
\hline
 2 & $(70,2^+)$ & $N 3/2^+$ & $N(1880) 1/2^+$ & $\Delta 3/2^+$ &   \\
  &  & $N(1860) 5/2^+$ & $N 3/2^+$ & $\Delta(2000) 5/2^+$ &   \\
  &  &  & $N(2000) 5/2^+$ & &   \\
  &  &  & $N(1990) 7/2^+$ & &   \\
\hline
 2 & $(20,1^+)$ & $N 1/2^+ $ &  & &   \\
 & & $N 3/2^+$ &  & &   \\
\hline
\hline
3 & $(70^*,1^-)$ & ${N(1895) 1/2^-}$ & $ N 1/2^- $ &$\Delta(1900) 1/2^-$  &  \\
  &  & $\bf{N(1875) 3/2^-}$ & $N(2150)3/2^-  $ &$\Delta(1940) 3/2^-$  &  \\
   &  &  & $N(2060) 5/2^- $ &  &  \\
\hline
\end{tabular}
\end{adjustbox}
\end{center}

\end{table}

It should be reminded that the allowed configurations for the $N_{h.o.}=3$ shell are \cite{fors}: 

\begin{equation}
(70^*,1^-), (56, 1^-), (70^{**},1^-),(20,1^-), (70^,2^-), (56, 3^-), (70,3^-),(20,3^-),
\end{equation}
but for simplicity in Table \ref{ho2} only the first configuration $(70^*,1^-)$ is considered.

The main drawback of the assignments shown in Table \ref{ho2} is  that in order to describe the extra negative parity N-states it is necessary to take into account also one of the $N_{h.o.}=3$ configurations, which, according to the h.o. model, should be systematically higher than the $N_{h.o.}=2$ states, while on the contrary  the experimental energies  are quite comparable with those of the $N_{h.o.}=2$ states. 

In the following we shall show that using the hypercentral Constituent Quark Model (hCQM) \cite{pl,chin} one can include all the negative parity states in the first three shells. This is possible thanks to the peculiar structure of the baryon spectrum provided by the hCQM and therefore we shall briefly recall its main features.

\section{The hypercentral Constituent Quark Model}

In CQMs it is customary to introduce  the Jacobi coordinates $\vec{\rho}$ and $\vec{\lambda}$,
\begin{equation}
\vec{\rho}~=~ \frac{1}{\sqrt{2}}(\vec{r}_1 - \vec{r}_2) ~,\\
~~~~\vec{\lambda}~=~\frac{1}{\sqrt{6}}(\vec{r}_1 + \vec{r}_2 - 2\vec{r}_3).  
\label{coord}
\end{equation}
In the hCQM, these coordinates are substituted with the hyperspherical coordinates \cite{baf}, which are given by the  angles ${\Omega}_{\rho}=({\theta}_{\rho},{\phi}_{\rho})$ 
and  ${\Omega}_{\lambda}=({\theta}_{\lambda},{\phi}_{\lambda})$ 
together with the hyperradius, $x$, and the hyperangle, $\xi$, defined as
\begin{equation}
x = \sqrt{\vec{\rho}^2 + \vec{\lambda}^2}, ~ ~ ~ ~ ~ ~ ~ ~ ~ ~ \xi = \arctan {\frac{\rho}{\lambda}}.
\end{equation}
Using these variables, the nonrelativistic kinetic energy operator, after having separated the c.m.\ motion,  can be written as
\begin{equation}
- \frac{\hbar^2}{2m} (\Delta_\rho + \Delta_\lambda) = - \frac{\hbar^2}{2m} ( \frac{\partial ^2}{\partial x^2}+\frac{5}{x} \frac{\partial}{\partial x}  + \frac{L^2(\Omega)}{x^2}).
 \end{equation}
The grand angular operator $L^2(\Omega)=L^2(\Omega_\rho, \Omega_\lambda,\xi)$ is the six-dimensional generalization
of the squared angular momentum operator.  Its eigenfunctions are the so called hyperspherical harmonics (h.h.) \cite{baf} $Y_{[\gamma]l_{\rho}l_{\lambda}}({\Omega}_{\rho},{\Omega}_{\lambda},\xi)$
\begin{equation}
L^2(\Omega_\rho, \Omega_\lambda,\xi)~Y_{[{\gamma}]l_{\rho}l_{\lambda}}({\Omega}_{\rho},{\Omega}_{\lambda},\xi)~=~-\gamma(\gamma+4) Y_{[{\gamma}]l_{\rho}l_{\lambda}}({\Omega}_{\rho},{\Omega}_{\lambda},\xi);
\label{grand}
 \end{equation}
the grand angular quantum number $\gamma$ is given by 
\begin{equation}
\gamma = 2 n + l_\rho + l_\lambda, 
 \end{equation}
where n is a nonnegative integer  and  $ l_\rho$, $l_\lambda$ are the angular momenta corresponding to the Jacobi coordinates of Eq.~(\ref{coord}). The h.h.\ describe the angular and hyperangular part of the three-quark wave function and are expressed by known products of standard spherical harmonics, trigonometric functions  and Jacobi polynomials \cite{baf}.

The starting point of the hCQM is the assumption that the quark interaction is hypercentral, that is it depends on the hyperradius x only \cite{pl,chin}
\begin{equation}
V_{3q}(\vec{\rho},\vec{\lambda})~=~V(x) ,
\end{equation}
therefore the space part of the three quark wave function $\psi_{space}$ of Eq. (\ref{3q_2})  is factorized
\begin{equation}
\psi_{space}~=~\psi_{3q}(\vec{\rho},\vec{\lambda})~=
\psi_{\gamma \nu}(x)~~
{Y}_{[{\gamma}]l_{\rho}l_{\lambda}}({\Omega}_{\rho},{\Omega}_{\lambda},\xi);
\label{psi}
\end{equation}
similarly to what happens in the standard three-dimensional problem. The hyperradial wave function $\psi_{\gamma \nu}(x)$  is labeled by the grand angular quantum
number $\gamma$ and  by the number of nodes $\nu$. The angular-hyperangular part of the 3q-state is completely described by the h.h.\ and is the same for any hypercentral potential. The dynamics is contained in the hyperradial  wave function $\psi_{\gamma \nu}(x)$, which is a solution of the hyperradial equation
\begin{equation}
[~\frac{{d}^2}{dx^2}+\frac{5}{x}~\frac{d}{dx}-\frac{\gamma(\gamma+4)}{x^2}]
~~\psi_{\gamma \nu}(x)
~~=~~-2m~[E-V_{3q}(x)]~~\psi_{\gamma \nu}(x).
\label{hyrad}
\end{equation}
 The  harmonic oscillator potential (h.o) often used in CQMs is exactly hypercentral, since 
 \begin{equation}
\sum_{i<j}~\frac{1}{2}~k~(\vec{r_i} - \vec{r_j})^2~=~\frac{3}{2}~k~x^2~=~V_{h.o}(x),
\label{ho3}
\end{equation}
the hyperradial equation (\ref{hyrad}) is solved analytically and the energy eigenvalues are given by
 \begin{equation}
E_{\gamma \nu}~=~ (3 +N_{h.o.}) \hbar \omega, ~~~~~~~~~~~N_{h.o.}~=~2 \nu + \gamma.
\label{en_ho}
\end{equation}

In the hCQM, the quark interaction is quite different from the h.o. one,  it has the form \cite{pl,chin}
\begin{equation}
V(x)~=~-\frac{\tau}{x} + \alpha x,
\label{h_pot}
\end{equation}
where $\tau$ and $\alpha$ are unknown parameters. This potential can be considered the hypercentral approximation of a Cornell-type quark interaction \cite{corn}, whose form can be reproduced by Lattice QCD calculations \cite{LQCD}.

Because of the limited space extension of the baryon states, the dominant part of the potential Eq. (\ref{h_pot})  \cite{sig,sig2} is given by the hyperCoulomb (hC) term
\begin{equation}
V_{hyc}(x)= -\frac{\tau}{x}. 
\label{hyc}
\end{equation}

Also the hC potential can be solved analytically \cite{sig,sig2,nc,hyc,breg} and the energy eigenvalues are given by
\begin{equation}
E_{n,\gamma}~=~-\frac{{{\tau}^2}m}{2(N_{hC}+\frac{5}{2})^2},~~~~~~~~~~~N_{hC}~=~ \nu+\gamma.
\label{en_hc}
\end{equation}

It should be stressed that the energies in Eqs. (\ref{en_ho}) and (\ref{en_hc}) have a quite different dependence on the quantum numbers $\gamma$ and $\nu$. In both cases the parity P of the states is
\begin{equation}
P~=~(-)^\gamma
\end{equation}
but
\begin{equation}
P_{h.o.}~=~(-)^{N_{h.o.}},     ~~~~~~~~~~~P_{hC}~\neq~(-)^{N_{hc}},
\end{equation}
that is the h.o. states in any given shell share the same parity, while in a hC shell states of both parities are present.

\begin{figure}[t]

\includegraphics[width=5in]{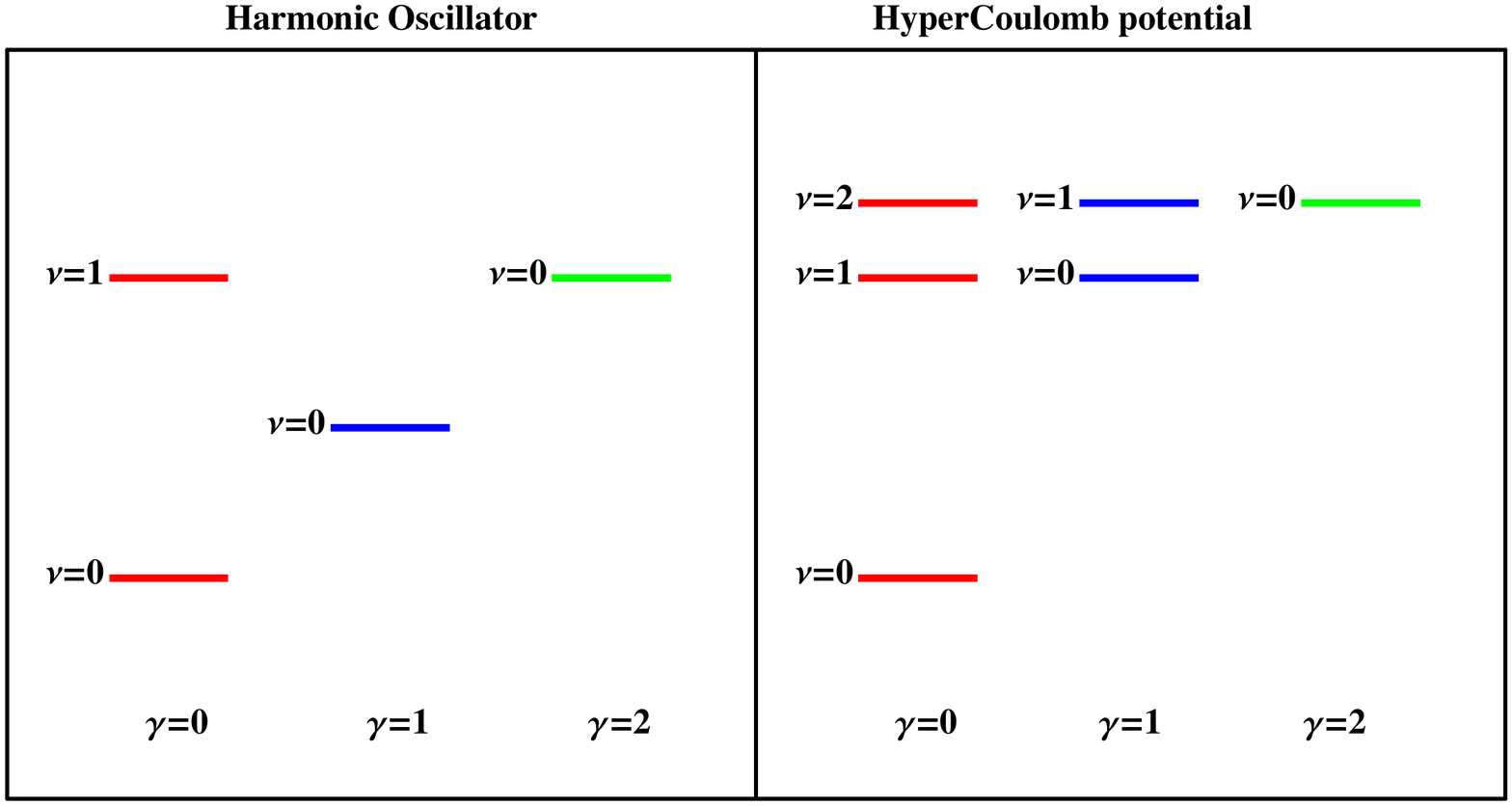}

\caption{ (Color online) Qualitative structure of  theoretical spectra for the h.o. (left) and for the hC potentials (right), up to the first three shells. The energy units are arbitrary and different for the two potentials.}
\label{ho-hc}
\end{figure}

The qualitative spectra of the h.o. and hC potentials, up to the first three shells, are shown in Fig. \ref{ho-hc}.

 For each level, the possible  three-quark states depend on the grand angular quantum number $\gamma$, the different values of $\nu$ representing the number of nodes of  the (hyper)radial excited states.

 \begin{eqnarray}
 \gamma=0 : & (56,0^+),  \\
 \gamma=1 : & (70,1^-),  \\
 \gamma=2 : & (70,0^+), (56,2^+),  (70,2^+),  (20,1^+).
 \end{eqnarray}

Because of the $\gamma+\nu$ dependence of the energies, the hyperradial excitations are degenerate with the negative parity states. This feature is of a x-dependent  potential, while any two-body potential leads to an ordering of the states similar to the h.o., that is, negative and positive parity states alternate each other. In the experimental spectrum the Roper resonance, which is the first radial excitation of the nucleon, is even lower than the negative parity resonances, a pattern which is more similar to the hC spectrum than to the h.o. one..

The hC allows to allocate  within the first three shells two negative parity levels, that is all the presently known negative parity resonances, without involving the higher shells, which have certainly a higher energy, while the new resonances have energies comparable to those of the $\gamma=2$ shell (see Table \ref{ho2}). The higher negative parity resonances are (hyper)radial excitations of the lower ones, since for them $\nu=1$.

Furthermore, since each shell of the hC contains a $(56,0^+)$ state,   there is also a second "Roper" resonance within the $N_{hc} \leq 2$ shells. This is the reason why in Table \ref{comp},  the hC has two more positive parity states with respect to the h.o. case.

\begin{figure}[h]

\includegraphics[width=4in]{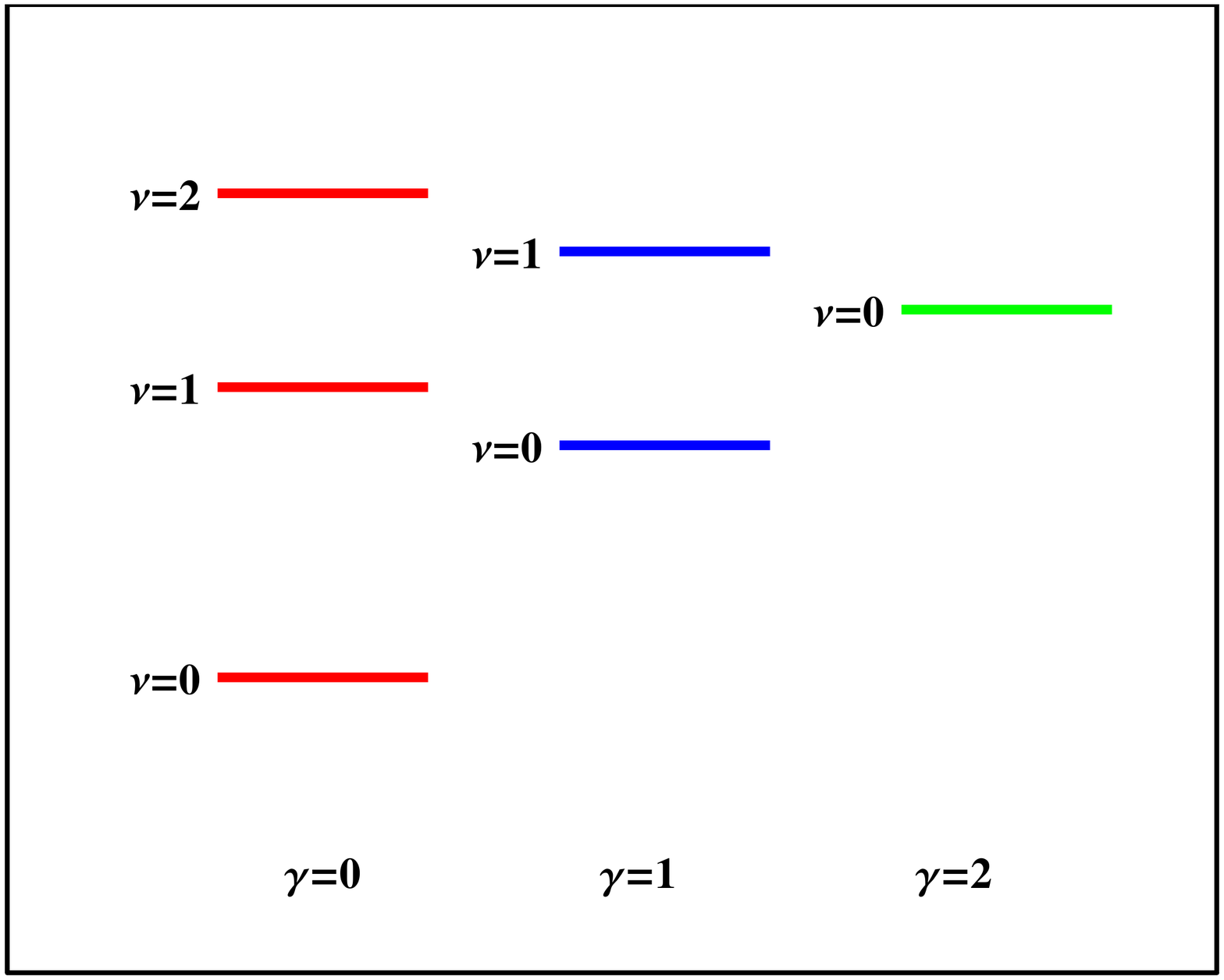}
\centering
\caption{ (Color online) Qualitative structure of  theoretical spectra for the hCQM potential of Eq.(\ref{h_pot}), up to the first three shells. The energy units are arbitrary.}
\label{conf}
\end{figure}

Coming back to the hCQM potential of Eq.(\ref{h_pot}),   a linear potential is added to the hC term, however the structure of the  spectrum given in the right part of Fig. \ref{ho-hc} is only slightly modified, as it is shown in Fig. \ref{conf}. This means that also with the complete hCQM potential all the known negative resonances can be described by the lower shells and a second hyper radial excitation of the nucleon is predicted.

\begin{figure}[t]

\includegraphics[width=5in]{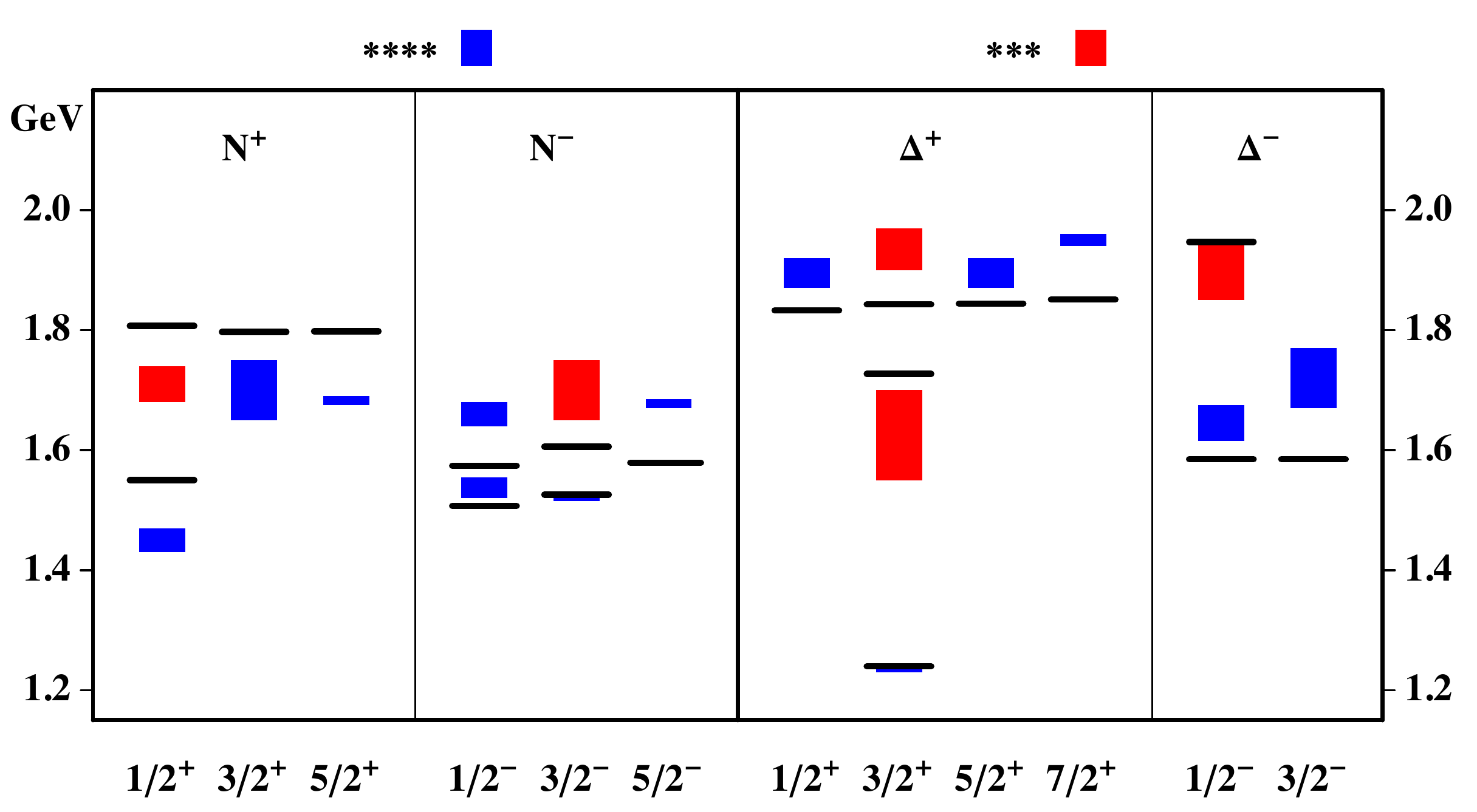}

\caption{ (Color online) The spectrum obtained using the hCQM hamiltonian of Eq.~(\ref{H_hCQM}). The free parameters are fitted  \cite{pl} to the experimental values of the 4* and 3* resonances reported in the $PDG94$ \cite{pdg94}.}
\label{fit}
\end{figure}

\begin{figure}[h] 

\includegraphics[width=5in]{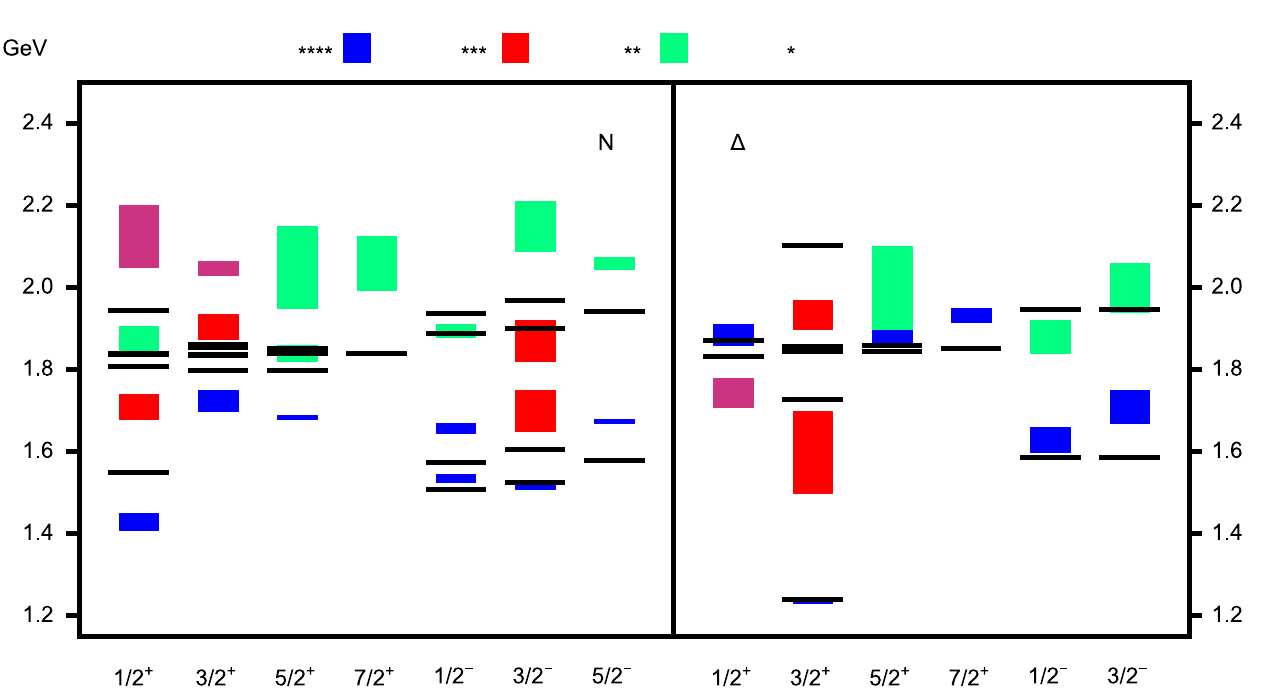}

\caption{ (Color online) The spectrum obtained using the hCQM hamiltonian of Eq.~(\ref{H_hCQM}), with the parameters fitted to the spectrum of Fig.\ref{fit}, in comparison with all the states reported in the PDG \cite{pdg14}. As explained in the text, the observed $2 *$ $N(2300) 1/2^+$, $N(2570)5/2^-$ and the three star $\Delta(1930) 5/2^-$ are omitted, the former two because their energy is too high and the latter because its quantum numbers are incompatible with the first three shells.}
\label{spec_all}
\end{figure}

The levels shown in Fig. \ref{conf} describe the average values of the $SU(6)$ configurations and therefore, in order to describe the splittings within the multiplets a violation term must be added. In the case of the hCQM, such violation of the $SU(6)$ symmetry is provided by the hyperfine interaction \cite{deru,ik}. The complete hCQM hamiltonian is then \cite{pl,chin}
\begin{equation}
H_{hCQM}~=~3m + \frac{\vec{p}_\rho^{~2}}{2m} + \frac{\vec{p}_\lambda^{~2}}{2m}-\frac{\tau}{x} + \alpha x + H_{hyp}.
\label{H_hCQM}
\end{equation}
where $\vec{p}_\rho$ and $\vec{p}_\lambda$ are the momenta conjugated to the Jacobi coordinates Eq.(\ref{coord}).

The quark mass is taken, as usual, $1/3$ of the nucleon one, there are then only three unknown parameters in the hamiltonian Eq. (\ref{H_hCQM}), namely $\alpha,\tau$ and the strength of the hyperfine interaction. These parameters have been fitted  \cite{pl} in order to reproduce the $4^*$ and $3^*$ resonances reported in the 1994 edition of the PDG \cite{pdg94}. Of course, the recent $3^*$ $N(1875)3/2^-$ state was not considered, but its energy can be now predicted. The fitted spectrum, taken form Ref. \cite{pl}, is reported in Fig. \ref{fit}.

The negative parity resonances are particularly well described, while the problem of the Roper is still present. However, it has been shown that in order to reproduce the position of the Roper it is necessary to consider also isospin dependent terms in the quark interaction \cite{iso}.

Having determined the free parameters of the model, it is possible to build the states for all the resonances of interest \cite{pl,sig,chin} and use them in order to predict various physical quantities  and compare the results with the experimental data. This has been done for the photocouplings \cite{aie}, the helicity amplitudes for the electroexcitation of the resonances \cite{aie2,sg} and for the elastic nucleon form factors \cite{mds,rap,ff_07,ff_10} (for a review see ref. \cite{chin}).

As already mentioned, in the hCQM there are two configurations more than those predicted by the h.o., namely one $(56,0^+)$ and one $(70,1^-)$. In the former there is a  further hyperradial excitation of both the nucleon and the $\Delta$, while in the latter one can insert further  five negative parity states, including the recently discovered $3^*$ $N(1875)3/2^-$. The number of states predicted by the hCQM in the first three shell is reported in the third column of Table \ref{comp}. It is seen that, if one considers also the $2^*$ states,  the number of negative parity states is very near to the observed one, while for the positive parity the theoretical resonances are still more than the experimental ones; the number of states predicted by the h.o. is in any case  lower than expected.  

The complete spectrum obtained with the hCQM is shown in Fig.\ref{spec_all} in comparison with all the states listed in the last edition of the PDG \cite{pdg14}, while the numerical values are reported in the second column of Tables \ref{hCQM_N} and \ref{hCQM_D}.

One should not forget that the hyperfine interaction produces a mixing of the $SU(6)$ configurations and the hCQM gives rise to definite superposition coefficients. The configurations can be labeled according to the notation introduced in Table \ref{ho} and the one corresponding to the maximum probability is reported in the third columns of Tables \ref{hCQM_N} and \ref{hCQM_D}. In some cases, the superposition amplitudes are quite comparable and the states are determined by a delicate balance of the mixed configurations.

Using the hCQM as a classification pattern and the results reported in Tables \ref{hCQM_N} and \ref{hCQM_D}, Table \ref{ho2} is substituted with Table \ref{hCQM}, where also the one star resonances are  given in square brackets.

The important feature of Table \ref{hCQM} is the inclusion of all the presently known negative parity states in the $N_{hC} \leq 2$ shells. But the presence of a second negative parity level has relevant consequences on the mixing of the hyperfine mixing, In fact, the $N 1/2^*$ and $N 3/2^-$ states are now a superposition of four SU(6) configurations: $(70,1^-,^28),(70^*,1^-,^28),(70,1^-,^48),(70^*,1^-,^48)$, leading to the interpretation of the $N(1650) 1/2^-$ and $N(1700) 3/2^-$ resonances as hyperradial excitations of the $N(1535) 1/2^-$ and $N(1520) 3/2^-$states, respectively. Furthermore, the new state $N(1875) 3/2^-$ is more suitably inserted in the $(70,1^-)$ level.

A further point of interest of Table \ref{hCQM} is the presence of a further hyper radial exception of both the nucleon and the $\Delta$, whose evidence is however absent, apart form the $1^* N(2100) 1/2^+$ state. 

The two states belonging to the $(70,1^+)$ configuration, with spin $1/2$ and $3/2$, respectively (see Table \ref{ho2}), are not affected by the hyperfine interaction unless higher shells are considered. They  are included in Tables  \ref{hCQM_N} and \ref{hCQM_D}, but being degenerate with other states are scarcely distinguishable in Fig. \ref{spec_all}.

In comparison with the states reported by the PDG \cite{pdg14}, in Table \ref{hCQM} many states are missing. Some of them have quantum numbers accessible to $N_{hC} \leq 2$ shells, namely the $2^*$ states $N(2300) 1/2^+$, $N(2570) 5/2^-$, which have a two high energy, and the $1^*$ state $\Delta(2150) 1/2^-$, which, if confirmed, should be certainly included in a higher shell, the lower ones being already full. 

All the other states not present in Table \ref{hCQM} have quantum numbers incompatible with those of the $N_{hC} \leq 2$ shells, as it happens for the $3^*$ $\Delta(1930)5/2^-$. However, also for the  states with higher energies, the degeneracy between positive and negative parity levels may be beneficial. For instance, the states $N(2220) 9/2^+$ and $N(2250) 9/2^-$ can be included in the same $N_{hC} = 3$ shell, while in the h.o. case they would belong to the $N_{h.o.} = 3$ and $N_{h.o.} = 4$ shells, respectively.

\begin{table}[h]
\caption[]{The energies of the N-states obtained with the hCQM \cite{pl} (second colunm), in comparison with the states reported  in the PDG \cite{pdg14} (fourth column) and their status (fifth column). The dominant $SU(6)$ configurations, obtained for each state by the hyperfine mixing, are given in the third column.}
\label{hCQM_N}
\vspace{15pt}
\center
\begin{tabular}{|c|c|l||c|c|}
\hline
 State & $M_{teor}  $ & ~~~$SU(6)$ &Baryon & Status \\
  & (MeV)  & ~~~conf.& PDG & \\
\hline
 $ N1/2^+$ & 938 & $(56,0^+) ~ ^28$ &  $N(939) 1/2^+$ & ****\\ 
  & 1550 & $(56^*,0^+)~ ^28$  &  $N(1440) 1/2^+$ & ****\\
 & 1808 & $(70,0^+) ~ ^28$  & $ N(1710) 1/2^+$ & ***\\
 & 1836 & $(20,1^+) ~ ^28$  & & \\
& 1839 & $(70,2^+)~ ^48$  & $ N(1880) 1/2^+$ & ** \\
 & 1943 & $(56^{**},0^+)~ ^28$ & $ N(2100) 1/2^+$ & * \\
\hline
 $ N 3/2^+$ & 1797 & $(56,2^+)~ ^28$  & $ N(1720) 3/2^+$ & **** \\
  & 1835 & $(70,2^+)~ ^48$ & &  \\
 & 1836 & $(20,1^+) ~ ^28$& & \\
  & 1853 & $(70,2^+) ~ ^28$ & $ N(1900) 3/2^+$ & *** \\
 & 1863 &  $(70,0^+) ~ ^48$   & $ N(2040) 3/2^+$ & *  \\
\hline
 $ N 5/2^+$ & 1798 & $(56,2^+)~ ^28$ & $ N(1680) 5/2^+$ & **** \\
  & 1844 & $(70,2^+)~ ^48$ & $ N(1860) 5/2^+$ & ** \\
  & 1851 &  $(70,2^+)~ ^48$& $ N(2000) 5/2^+$ & ** \\
 \hline
  $N 7/2^+$ & 1840 &  $(70,2^+)~ ^48$&   $ N(1990) 7/2^+$ & ** \\
\hline
 \hline
 $ N 1/2-$ & 1507 & $(70,1^-)~ ^28$ &  $ N(1535) 1/2^-$ & ****\\
  & 1574 & $(70^*,1^-)~ ^28$ & $ N(1650) 1/2^-$ & **** \\
   & 1887 & $(70,1^-)~ ^48$ & $ N(1895) 1/2^-$ & ** \\
  & 1937& $(70^*,1^-)~ ^48$ & & \\
\hline
 $ N 3/2-$ & 1525 & $(70,1^-)~ ^28$ & $ N(1520) 3/2^-$ & **** \\
  & 1606 & $(70^*,1^-)~ ^28$ & $ N(1700) 3/2^-$ & *** \\
  & 1899 & $(70,1^-)~ ^48$ & $ N(1875) 3/2^-$ & *** \\
 & 1969 & $(70^*,1^-)~ ^48$ & $ N(2150) 3/2^-$ & ** \\
 \hline
 \hline
 $ N 5/2-$ & 1579 & $(70,1^-)~ ^48$ & $ N(1675) 5/2^-$ & **** \\
  & 1942 & $(70^*,1^-)~ ^48$ & $ N(2060) 5/2^-$ & ** \\

 \hline

\end{tabular}

\end{table}

The theoretical spectrum of Fig. \ref{spec_all} is somewhat compressed towards the lower energies, an effect which is mainly due to hC component of the interaction. Furthermore, the hyperfine interaction in the case of the hCQM, having  fixed its strength in order to reproduce the Nucleon-$\Delta$ splitting, gives in general smaller effects than the IK model \cite{ik}. In any case, as we have already noted, in order to get  a better description of the spectrum, one needs also isospin dependent terms \cite{iso}.

\begin{table}[t]
\caption[]{The same as in Table \ref{hCQM_N}, but for the $\Delta$ states. }
\label{hCQM_D}
\vspace{15pt}
\center
\begin{tabular}{|c|c|c||c|c|}
\hline
 State & $M_{teor}  $ &  ~~~$SU(6)$ & Baryon & Status \\
  & (MeV)   &  ~~~conf.  & PDG & \\
\hline
 $ \Delta1/2^+$ & 1832 &  $(70,0^+) ~ ^210$  &$\Delta(1750) 1/2^+$ & *\\ 
  & 1871 & $(56,2^+) ~ ^410$ & $\Delta(1910) 1/2^+$ & ****\\
 \hline
 $ \Delta 3/2^+$ & 1240 & $(56,0^+) ~ ^410$ & $ \Delta(1232) 3/2^+$ & **** \\
  & 1727 & $(56^*,0^+) ~ ^410$ & $ \Delta(1600) 3/2^+$ & ***  \\
  & 1843 & $(70,2^+) ~ ^210$ & &  \\
 & 1856 & $(56,2^+) ~ ^410$ &  $ \Delta(1920) 3/2^+$ & *** \\
  & 2103 & $(56^{**},0^+) ~ ^410$ &  & \\
\hline
 $ \Delta 5/2^+$ & 1844 & $(56,2^+) ~ ^410$ &  $  \Delta(1905) 5/2^+$ & **** \\
  & 1859 & $(70,2^+) ~ ^210$ &$  \Delta(2000) 5/2^+$ & ** \\
 \hline
  $\Delta 7/2^+$ & 1851 & $(56,2^+) ~ ^410$ &   $  \Delta(1950) 7/2^+$ & **** \\
\hline
 \hline
 $ \Delta 1/2-$ & 1584 & $(70,1^-)~ ^210$ &  $  \Delta(1620) 1/2^-$ & ****\\
  & 1947 & $(70^*,1^-)~ ^210$ &  $  \Delta(1920) 1/2^-$ & ** \\
\hline
 $ \Delta 3/2-$ & 1584 & $(70,1^-)~ ^210$ & $  \Delta(1700) 3/2^-$ & **** \\
  & 1947 & $(70^*,1^-)~ ^210$ & $  \Delta(1940) 3/2^-$ & ** \\
 \hline

 \hline

\end{tabular}
\end{table}

\begin{table}[h]
\caption[]{The non strange three-quark states belonging to  the hCQM \cite{pl} shells up to $N_{hC}=2$. The notation is the same as in Table \ref{ho2}. }
\label{hCQM}
\vspace{15pt}
\begin{center}
\begin{adjustbox}{max width=\textwidth}
\begin{tabular}{|c|c|c|c|c|c|}
\hline
$N_{hC}$ & $(d,L^P)$ & $^2\underline{8}$ & $^4\underline{8}$ & $^2\underline{10}$ & $^4\underline{10}$ \\
\hline
\hline
 0 & $(56,0^+)$ & $\bf{N(939) 1/2^+}$ &  &  &  $\bf{\Delta(1232) 3/2^+}$ \\
\hline
\hline
 1 & $(56^*,0^+)$ & $\bf{N(1440) 1/2^+}$ &  &  &  $\bf{\Delta(1600) 3/2^+}$ \\
\hline
 1 & $(70,1^-)$ & $\bf{N(1535) 1/2^-}$ & $N(1895) 1/2^- $ &$\bf{\Delta(1620)}1/2^-$  &  \\
  &  & $\bf{N(1520) 3/2^-}$ & $\bf{N(1875) 3/2^-} $ &$\bf{\Delta(1700)} 3/2^-$  &  \\
   &  &  & $\bf{N(1675) 5/2^-} $ &  &  \\
\hline
\hline
2 & $(70^*,1^-)$ & $\bf{N(1650) 1/2^-}$ & $ N 1/2^- $ &$\Delta(1900) 1/2^-$  &  \\
  &  & $\bf{N(1700) 3/2^-}$ & $N(2150)3/2^-  $ &$\Delta(1940) 3/2^-$  &  \\
   &  &  & $N(2060) 5/2^- $ &  &  \\
\hline
  2 & $(70,0^+)$ & $\bf{N(1710) 1/2+}$ & $ [ N(2040) 3/2^+] $  & $\Delta 1/2^+$ &   \\
\hline
 2 & $(56,2^+)$ & $\bf{N(1720) 3/2^+}$ &  &  & $\bf{\Delta(1910) 1/2^+}$  \\
  &  & $\bf{N(1680) 5/2^+ }$ &  &  & $\bf{\Delta(1920) 3/2^+} $  \\
  &  &  &  &  & $\bf{\Delta(1905) 5/2^+}$  \\
  &  &  &  &  & $\bf{\Delta(1950) 7/2^+}$  \\
\hline
 2 & $(70,2^+)$ &$\bf{N(1900) 3/2^+}$ & $N(1880) 1/2^+$ & $\Delta 3/2^+$ &   \\
  &  & $N(1860) 5/2^+$ & $N 3/2^+$ & $\Delta(2000) 5/2^+$ &   \\
  &  &  & $N(2000) 5/2^+$ & &   \\
  &  &  & $N(1990) 7/2^+$ & &   \\
\hline
 2 & $(20,1^+)$ & $N 1/2^+ $ &  & &   \\
 & & $N 3/2^+$ &  & &   \\
\hline
2 & $(56^{**},0^+)$ & $[N(2100) 1/2^+] $ &  &  &  $\Delta 3/2^+$ \\
\hline

\end{tabular}
\end{adjustbox}
\end{center}
\end{table}

\section{Conclusions}

The predictions of baryon states by any CQM have been always affected by the problem of the missing resonances. It is interesting to note that the recent editions of the PG \cite{pdg12,pdg14} have reported various new states, with different levels of certainty (stars) and  therefore the distance between the number of theoretical and experimental states is  reduced. 

 Looking at  Table \ref{comp}, one sees that the number of the observed positive parity N and $\Delta$ states ranges from 12 for 4-and 3-* resonances up to 20 for states with any number of stars. If one refers, as usual, to the h.o., the positive parity states should be to 23. The problem of the missing resonances is still present, but not too serious. 
 
 The situation is completely different for the negative parity states. Thanks to the new entry $N(1875) 3/2^-$ \cite{pdg12,pdg14}, the observed 4-and 3-* resonances are 8 while the h.o. ones are 7. The total number of PDG states, irrespective of the star assignments, is 13, taking into account also of the $\Delta(2150) 1/2^-$ state, not included in Table \ref{comp}. Therefore, the theoretical pattern based on the h.o. is not able to account for  the observed spectrum. 
 
On the other hand, in the hCQM \cite{pl,chin}, because of its dependence on the hyperradius, the level pattern is quite different from the h.o. one, as it is shown in Figs. \ref{ho-hc} and \ref{conf}. In fact, in each shell, both positive and negative parity states are present. This means that in the first level there is the usual hyper radial excitation of the nucleon together with 7 negative parity states. In the third shell, besides the states predicted by the h.o., there is a second negative parity level in which seven further states can be accommodated and can be interpreted as hyperradial excitations of the ground negative parity resonances. This peculiar feature of the hCQM is therefore beneficial in two respects: the energy of the Roper resonance is comparable with that of the negative states and the newly observed negative  states can be more naturally included in the third shell and not in the fourth one, which is for sure too high in energy. 

The hCQM provides then a better basis for the description of the spectrum. The price to pay is the prediction of a further hyper radial excitation of the nucleon, for which possible candidates are the 1* $N(2100) 1/2^+$ or the 2* $N(2300) 1/2^+$ states, and also of the  $\Delta$, or which there seems to be presently no candidate. Furthermore, there is  an extra negative parity state (the 1* $\Delta(2150) 1/2^-$), which cannot be fitted into the scheme; however it is in the higher part of the energy and it presumably belongs to the $N_{hC}=3$ shell. 

A particular problem is provided by the $3^*$ $\Delta(1930)5/2^-$, whose quantum  numbers, as already mentioned, cannot be obtained with states belonging to the first shells, even in the hCQM, where two $(70,1^-)$ levels are available. It should be  a member of the $N_{hC}=3$ shell.

The emphasis is here  mainly devoted  to  qualitative aspects, first of all to the possibility of  including in the theoretical description also the recently discovered resonances and the  agreement of theoretical spectrum with the observed one. In this respect, there are interesting similarities between the spectra obtained with  the hCQM and with  LQCD calculations \cite{edw}: in fact, in both cases,  the medium-high energy region is more populated with  respect to the h.o. case and there is a compression in the lower energy part.

As the energy increases, the number of predicted states becomes very large, in strong contrast with the few observed states. However, one should not forget that {\bf any} CQM predicts baryon states with zero width, since no coupling with the continuum is present, whereas resonances have a large width and, if their number increases, there is a strong overlapping, which presumably prevents from their identification, apart maybe in the case of high spin values.

In order to take into account the coupling with the continuum,  one should include in the CQM in a consistent way the quark-antiquark pair creation mechanisms, that is, one should proceed to the unquenching of the model.

Such unquenching has been performed some time ago in the meson sector \cite{gi}, but recently it has been extended with success also to the case of baryons \cite{sb1,bs,sb2}.

The unquenching of the CQM has many important consequences on the description of baryons. The baryons are  no longer a simple three quark states, but also higher Fock components must be introduced. It has been shown \cite{bs} that in this way the old good results on the magnetic moments are unaffected. Probably it happens here also what it has been shown in the meson case \cite{gi}, that is the inclusion of quark-antiquark pairs leads to a renormalization of the string constant. 

Nevertheless, the inclusion of such mechanisms is beneficial for the description of the low $Q^2$ behavior of the helicity amplitudes for the electroexcitation of the baryon resonances: in fact, there is a consensus  that the lack of strength at low $Q^2$ predicted by all CQMs may be due just to the missing quark-antiquark pair effects \cite{aie2,sg,chin}. Furthermore, having at disposal an unquenched CQM, one can calculate in a consistent way the electroproduction of mesons and eventually describe exotic states, such as pentaquarks (for a classification of all possible qqqq\={q} states see Ref. \cite{bgs}). Also in the case of the elastic nucleon form factors, the quark-antiquark pair effects are expected to be important \cite{rap,ff_07,ff_10}.

In any case, the unquenching will make CQMs  suitable for the description of a wider class of  baryon properties.

\end{document}